\documentclass[12pt,preprint]{aastex} %original
\newcommand{\cerenkov}{{\v C}erenkov}

\newcommand{\gam}{$\gamma$}
\def\3EG{{3EG J1746-2851}}
\def\p0{{$\pi^0$}}
\def\1018{{$10^{18}$}}

\newcommand{\be}{\begin{equation}}
\newcommand{\ee}{\end{equation}}
\newcommand{\bea}{\begin{eqnarray}}
\newcommand{\eea}{\end{eqnarray}}
\newcommand{\bmu}{\begin{multline}}
\newcommand{\emu}{\end{multline}}

\def\numu{{$\nu_\mu  $}}
\def\nue{{$\nu_e  $}}
\def\nutau{{$\nu_\tau  $}}

\def\simlt{\lower.5ex\hbox{$\; \buildrel < \over \sim \;$}}
\def\simgt{\lower.5ex\hbox{$\; \buildrel > \over \sim \;$}}

\def\gcm3{{\rm\,g\,cm^{-3}}}
\def\ncm3{{\rm\,cm^{-3}}}

\def\>{$>$}
\def\<{$<$}

\lefthead{Crocker et~al.}                      
\righthead{Galactic Center Neutrinos}              

\received{}
\begin{document}
%\title{\bf Neutrinos from the Galactic Center in the Light of HESS}
\title{\bf Neutrinos from the Galactic Center in the Light of its Gamma-Ray
Detection at TeV Energy}
\author{Roland M. Crocker\altaffilmark{1}$^,$ \altaffilmark{2},
Fulvio Melia\altaffilmark{3}, 
and Raymond R. Volkas\altaffilmark{2}}

%\author{Roland M. Crocker,$^{1,2}$ Fulvio Melia,$^{3}$ 
%\\and Raymond R. Volkas$^2$}
%\affil{$^1$ Harvard-Smithsonian Center for Astrophysics\\
%60 Garden St.,
%Cambridge MA 02138\\
%rcrocker@cfa.harvard.edu}
\altaffiltext{1}{Harvard-Smithsonian Center for Astrophysics,
60 Garden St.,
Cambridge MA 02138;
rcrocker@cfa.harvard.edu}
%\affil{$^2$Research Centre for High Energy Physics, 
%School of Physics,
%\\The University of Melbourne,
%3010 Australia\\
%r.crocker, r.volkas@physics.unimelb.edu.au}
\altaffiltext{2}{Research Centre for High Energy Physics, 
School of Physics,
The University of Melbourne,
3010 Australia;
r.crocker, r.volkas@physics.unimelb.edu.au}
%\affil{$^{3}$Physics Department and Steward Observatory, 
%\\The University of Arizona, Tucson, AZ 85721\\
%melia@physics.arizona.edu}
\altaffiltext{3}{Physics Department and Steward Observatory, 
The University of Arizona, Tucson, AZ 85721;
melia@physics.arizona.edu}

\begin{abstract}
We re-evaluate the event rate expected in 
km$^3$-scale detectors for neutrinos from the direction of the Galactic
Center (GC) in light of recent spectral measurements obtained by the HESS
instrument for $\sim$TeV \gam -radiation from this direction. 
In the most plausible scenario the re-evaluated event
rate is smaller than that previously calculated---and here 
re-calculated---on the basis of EGRET data.
However, 
the GC TeV \gam -ray detections by the Whipple, CANGAROO, and HESS instruments, 
together with the strong indications
for an  overabundance of cosmic rays 
coming from the GC at EeV energies, 
strengthen the expectation for a detectable, TeV-PeV GC neutrino signal
from proton-proton interactions in that region.
If the TeV gamma-ray--EeV cosmic ray anisotropy connection is correct, 
this signal will be 
detectable within a year and half for km$^3$-scale 
neutrino detectors in the Northern Hemisphere at super-TeV energies and, 
significantly, 
should also be detectable in 1.6 years
by the South Polar IceCube detector at 
energies $\gtrsim 10^{14}$ eV.
The GC neutrino signal should also produce
a detectable 
signal from neutrino showering and
resonant $W^-$ production by $\overline{\nu}_e$'s 
in the volume of a km$^3$-scale detector.
\end{abstract}
\keywords{cosmic-rays --- elementary particles --- 
 Galaxy: center --- neutrinos ---
radiation mechanisms: nonthermal --- supernova remnants}

\section{Introduction}
Several of us \citep{Crocker2000,Crocker2002,Blasi2004} have previously calculated 
the flux of neutrinos expected from the Galactic center (GC) based on the $\pi^0$-decay 
EGRET \gam -ray signal \citep{Fatuzzo2003,Melia1998}.
%Similar calculations have been made by other authors \citep{Alvarez2002}.  
However, 
a new calculation is now warranted in light of (i) new $\sim$TeV $\gamma$-ray observations
of the GC with the Whipple and HESS air {\v C}erenkov telescopes 
\citep{Aharonian2004,Kosack2004}, (ii)
analyses that indicate that the EGRET GeV source is offset from the actual GC
\citep{Hooper2002,Pohl2004},
and (iii) recent theoretical progress in
understanding the 
totality of high-energy, GC astroparticle data \citep{Crocker2004}.
In particular, both the 
extremely high-energy (EHE) GC
cosmic ray anisotropy %\citep{Hayashida1999,Bellido2001}, 
(Hayashida et~al. 1999a; Bellido et~al. 2001)
and the GC \gam -ray signals can be ascribed, respectively,
to neutrons and neutral pions created 
%in the GC region
by the collisions of protons
from
the same shock-accelerated, GC population 
%(with spectral index near 2.2)
with ambient protons.

In this picture, neutrinos
too will be created  as the result of the decay of the charged pions 
arising inevitably
from the
same {\it pp} interactions. 
In fact, we can normalize the expected neutrino flux to the
\gam -ray and (putative) neutron fluxes
because of the common origin of all these particle species.
On this basis,
in this {\it Letter}, we re-calculate both the flux of high-energy 
neutrinos from the GC and the resulting event rates in the large 
scale neutrino telescopes\footnote{The implications of the putative GC neutron beam
for the GC neutrino flux have also been examined
by %\citet{Anchordoqui2004}.
Anchordoqui et~al. (2004a).
These authors employ quite different particle physics to explain the
neutrons, however, and their models do not address the GC \gam -ray signal.
}. Because of the EHE neutron connection, we expect the GC 
neutrino flux to extend to much higher energies than previously anticipated, 
meaning that it should now also be detectable through the resonant interactions of 
GC $\overline{\nu}_e$'s with electrons in the  volume of a km$^3$-scale detector.

\section{GC \gam-ray Data: Evidence for Hadronic Acceleration}

The GC has been detected in \gam -rays by the EGRET instrument aboard the
Compton Gamma-ray Observatory \citep{Mayer-Hasselwander1998}, CANGAROO 
%\citep{Tsuchiya2004}
(Tsuchiya et al.~ 2004), Whipple \citep{Kosack2004}
and HESS \citep{Aharonian2004}.  The latter three cover a similar energy
range, $\sim 10^{-1}-10$ TeV, while EGRET has lower energy data ($\sim 10^{-5} - 10^{-2}$ TeV). The Whipple result, while of
limited statistical significance, shows a constant flux over a decade,
and a flux consistent with the HESS result (K. Kosack 2005, private communication), 
extending to energies of at
least 2-3 TeV.
Because HESS has by far the best angular resolution and most detailed
spectral results and these results are consistent with 
%the earlier
the 1995-2000
Whipple detections, 
%over the 1995-2000 period,
we hereafter choose to employ the HESS spectrum, rather than that 
of CANGAROO in our analysis. We now %briefly
review why these data point to a hadronic origin.

The EGRET 
spectrum exhibits a clear break at $\sim 1$ GeV
which
can be explained by neutral pion decays generated in
collisions between relativistic and ambient protons. 
Such decays
produce a broad 
$\gamma$-ray feature that mirrors all but the lowest energy EGRET datum
which is, instead, explained self-consistently (in steady state) as
resulting from the $\gamma$-ray emission via bremsstrahlung and Compton scattering
of the charged leptons resulting from the decay of the charged
pions also produced
in these collisions.
Further increasing our confidence in this picture, Fatuzzo and Melia (2004) have shown 
that when the same 
charged secondary lepton population
is placed in a magnetic field sufficient to accelerate 
protons up to the required $\sim 10^{19}$ eV, it synchrotron radiates
with an emissivity also in agreement with radio observations.

The TeV observations mentioned above all also lend crucial support 
to the notion
that high-energy hadronic processes are taking place at the GC
because, as argued by Crocker et~al. (2004)
%our previous work 
%\citet{Crocker2004},
these data then reliably {\it predict} the GC neutron flux
apparently uncovered in the EHE cosmic ray observations.
Non-hadronic origins of the HESS signal are possible, though,
and have recently been discussed by \citet{Aharonian2004b}.

Clouding these waters, however, 
the entirety of GC, astroparticle data can not be fit with a model
involving proton shock acceleration and subsequent collision with ambient protons
{\it at a single source} \citep{Crocker2004}: 
%This is evidenced by the fact that
though a very good fit is possible to the EGRET+EHECR data, such a fit
over-predicts the $\gamma$-ray flux from the GC at HESS energies by a factor of
$\sim$20. 

%Taking all data at face value, there are two reasonable resolutions to this disagreement:
There are two reasonable resolutions of this:
(i) the TeV flux is  attenuated in propagation 
in which case the neutrino flux should be normalized to the combination of the
unattenuated EGRET data and the EHECR data; 
(ii) there are two {\it effective} GC sources, 
in which case the EGRET source must cut-off well below a TeV in order
not to pollute the HESS signal. In support of (ii) note that 
(a) the EGRET data themselves can, indeed, be interpretted as 
indicating just such a cut-off and (b) 
in analyses by
Hooper and Dingus (2002)
and Pohl (2004) an offset between
the EGRET and TeV sources is evident.
Finally, 
it is also possible that future observations
%---from the Auger extensive air shower
%array \citep{Bertou2002} in particular---
will fail to confirm the existence of the 
GC cosmic-ray anisotropy.
Alternatively, it might be established---despite all current 
contra-indications---that 
the GC \gam -ray and EHECR signals are   unrelated.
Given these possibilities, we should also consider the implications for the GC neutrino flux
of the HESS GC results {\it alone}.
We shall compute the GC neutrino flux for all these three 
cases (labeled EGRET+EHECR, HESS+EHECR, and HESS ALONE) below.
%\null\vskip-0.5in\null

\section{Neutrino Fluxes}

For the purposes of this study, we have generalized the
standard technique based upon ``spectrum weighted moments''
\citep{Gaisser1990}
to allow for 
%A standard technique based upon ``spectrum weighted moments (SWM)'' can be used
%to inter-relate the $\gamma$-ray, $n$ and $\nu$ emissitivites from 
%the GC.\citep{Crocker2000,Crocker2002,Crocker2004,Blasi2004}  
%In essence,
%the emissivity of species $x$ as a function of energy is equal to
%the product of the SWM for $x$ and the number per unit time per unit energy
%density of parent protons interacting with ambient protons, provided
%spectral index $\gamma$ (resulting in a daughter particle population with the same index)
%and (ii) the collisions take place in a center of mass energy regime for which scaling is obeyed. 
%
%For the present study we have generalized the
%SWM technology to allow for 
(i) an exponential cut-off in the parent particle
spectrum (which produces, to a good approximation, a mirroring
exponential cut-off in the daughter particle population, though with a reduced
cut-off energy), (ii) a scaling-violating growth of the total cross-section over the large
energy ranges separating the different sorts of data,
and (iii) various propagation effects.

With these generalizations in place,
we can relate
the fluxes of the various particle species (assuming that all the detected particles
are created in the same interaction process). 
In particular, in 
%our previous work 
(Crocker et~al.~2004), 
we determined the theoretical
relation between the $\gamma$-ray and neutron fluxes of the GC---at the vastly different 
energy scales  of $\sim$MeV/GeV and $\sim$10$^{18}$ eV---and were then able to perform 
simultaneous fits (in spectral index and $\gamma$-ray differential flux at some
normalizing energy) to the EGRET+EHECR data and the HESS+EHECR
data.
This required that we account
for the propagation effect of neutron decay-in-flight.

As mentioned above, the fit to the EGRET+EHECR data
only makes sense given  another propagation effect is operating:
attenuation of 
the TeV gamma rays. A  possible attenuation
mechanism  is $\gamma \gamma$ pair production
on the background NIR photons emitted by the circumnuclear disk at the GC.
It seems difficult, however, to arrange  for a
column density of NIR photons from the GC
sufficient to produce the required
attenuation. Further, 
were such attenuation to take place, the most natural expectation
would then be that the resulting spectrum is distorted away
from the initial (flat) power law.
The lack of any such distortion in the observed spectrum
and the relatively small column density of NIR photons together imply
that
%, though 
%normalizing the neutrino flux to the EGRET+EHECR data 
%generates the largest expected signal, 
this scenario seems unlikely
(see %\citet{Crocker2004} 
Crocker et~al. 2004
for more detail here).
We examine this possibility in our analysis, then, only in the
spirit that it provides something like an upper limit to 
the flux of neutrinos from the GC (due to conventional physics).

In the more compelling HESS+EHECR scenario, the neutrino
fluxes are due, in principle, to two (effective) sources and should be normalized to the 
cut-off
$\gamma$-ray flux measured by EGRET and the combination of the HESS and EHE CR data.
In practice, however, because the cut-off energy for the EGRET source
\gam -rays must be in the 100 GeV energy range, the neutrino spectrum of this 
source will be similarly cut-off rendering it invisible to km$^3$-scale
detectors against the atmospheric neutrino background (given
reasonable values for detector angular resolution).

Finally, as presaged above,
we examine for completeness the consequences of normalizing the GC 
neutrino flux to the HESS data by themselves. 
We include two cases here. The first is where the cut-off in the
photon spectrum is taken to be at $\sim$10$^{17}$ eV. This case is
numerically identical to a pure power-law
fit to the HESS data which, it should be noted, show no
direct evidence for a cut-off. Second, {\it in order to
arrive at a strict lower limit} to the GC flux
we examine the case of $E_\gamma^{cut} = 10^{13}$ eV. This is the
approximate {\it minimum} cut-off energy consistent with the HESS data.

Now, employing these fitted normalizations and spectral indices, we wish to calculate
the muon and electron type neutrino fluxes  coming
from the GC direction.  This calculation must account 
for a further propagation effect, viz. in-vacuum neutrino oscillations.
Given the distance and energy scales involved these will be totally averaged out 
(unless sub-dominant, long-wavelength oscillation modes operate in nature; 
Crocker et~al. 2002),
implying flavor ratios close to 
$\nu_e : \nu_\mu : \nu_\tau = 1 : 1 : 1$ at the Earth.
%\citep{Learned1995,Beacom2003}.
With these inputs we find 
$\Phi_{\nu_e}[E_\nu] \simeq \Phi_{\nu_\mu}[E_\nu] = \Phi_{\nu_\tau}[E_\nu] \equiv \Phi_{\nu}[E_\nu]$
%\be
%\qquad\Phi_{\nu_e}[E_\nu] \simeq \Phi_{\nu_\mu}[E_\nu] = \Phi_{\nu_\tau}[E_\nu] \equiv \Phi_{\nu}[E_\nu]
%\ee
and the following neutrino fluxes:
\bea
\hbox{EGRET+EHECR:}\qquad\Phi_{\nu}[E_\nu] 
&=& 9.0 \times 10^{-11} 
\left(\frac{E_\nu}{\mathrm{TeV}}\right)^{-2.22}  
\ \mathrm{cm}^{-2} \ \mathrm{s}^{-1} \ \mathrm{TeV}^{-1} \;
\eea
\bea
\hbox{HESS+EHECR:}\quad\qquad\,\Phi_{\nu}[E_\nu] 
&=& 1.3 \times 10^{-12} 
\left(\frac{E_\nu}{\mathrm{TeV}}\right)^{-2.00}  
\ \mathrm{cm}^{-2} \ \mathrm{s}^{-1} \ \mathrm{TeV}^{-1} \;
\eea
\bea
\hbox{HESS:}\qquad\,\Phi_{\nu}[E_\nu] 
&=& 1.2 \times 10^{-12} 
\left(\frac{E_\nu}{\mathrm{TeV}}\right)^{-2.23}  
\exp\left[\frac{- E_\nu}{E_\nu^\mathrm{cut}}\right]
\ \mathrm{cm}^{-2} \ \mathrm{s}^{-1} \ \mathrm{TeV}^{-1} \; ,
\eea
where ${E_\nu^{\mathrm{cut}}}
\in \{10^{17
} \ \textrm{eV},10^{13} \ \textrm{eV}\}$.
The neutrino fluxes above are plotted in 
Figure \ref{figl} which also shows backgrounds to \numu \ CC events   
in a km$^3$ Mediterranean detector and to IceCube, labeled,
``Atm \numu'' and ``Atm $\mu$'', respectively.
Here the former corresponds to
the atmospheric
\numu \ flux inside a solid angle encircling the GC direction
defined by the predicted 
angular resolution of the ANTARES neutrino telescope and the 
latter is the
atmospheric muon flux, at  a fiducial 1.6 km depth in the ice,
inside the predicted IceCube angular resolution
{\it weighted by the reciprocal of the (energy-dependent) neutrino
detection probability}. 
%\null\vskip-0.5in\null
\section{Event Rates}
From these fluxes we now calculate the event rates in astrophysical 
neutrino detectors. In general, the yearly event rate in such devices will be
given by 
\be
\mathrm N_{\mathrm year} 
= \int_{E_\nu^{min}} d E_\nu \int_0^{\mathrm year} d t \;
\hbox{Area}[E_\nu,\theta(t)] \; \Phi[E_\nu] 
 \times {P_{\mathrm{detect}}} [E_\nu] \; \hbox{Attn}\left[E_\nu,\theta(t)\right] \, .
\label{eqn_dtcnpr}
\ee
Here $\hbox{Area}[E_\nu,\theta]$ is the energy- and nadir-angle-dependent effective
(muon) area of the detector and
${P_{\mathrm{detect}}} [E_\nu] $ is the probability
that a neutrino will interact sufficiently close to the detector volume
that a detectable signal (muon track, electromagnetic or hadronic shower, etc.)
is created. The $\hbox{Attn}\left[E_\nu,\theta\right]$ function accounts
for neutrino interactions in the Earth's volume before the detector is
reached. For this function we employ a parameterization of the results
in {Naumov} and {Perrone} (1999).
%\citet{Naumov1999}.

In Equation (\ref{eqn_dtcnpr}), both $\hbox{Area}[E_\nu,\theta]$ 
and ${P_{\mathrm{detect}}} [E_\nu] $
depend on the detector and the neutrino interaction
process generating the signal. For CC
interactions
of \numu 's leading to muon tracks through the volume of 
a H$_2$O-based neutrino telescope, 
we 
employ the detection
probability presented by 
%\citet{Halzen2002}.
{Halzen} \& {Hooper} (2002).
Such detectors include,
buried in the ice at the South Pole,
(the currently-operating) AMANDA \citep{Ahrens2004b} and
(AMANDA's under-construction, km$^3$-scale replacement)
IceCube \citep{Ahrens2004}, and, 
in the deep Mediterranean,
the prototype-stage, $\lesssim$ 0.1 km$^2$ area ANTARES \citep{Korolkova2004}
%, 
%NESTOR \citep{Tsirigotis2004}, and NEMO \citep{Capone2003}
or a future, km$^3$-scale upgrade of this device.
%\footnote{Details of a proposal
%for such a detector---to be called ``Km3Net"---can be found at 
%http://www.km3net.org.}. 

For the showers created by the CC interactions of
\nue 's or \nutau 's (the latter without visible $\tau$ track)
and the neutral current interactions of all flavors
we employ the  event rate estimation set out by Beacom et~al. 
(2003).
For the event rate due to the 
%\citet{Glashow1960} 
Glashow (1960) process---i.e., the 
resonant, s-channel creation of W$^-$'s in $\overline{\nu}_e$ e$^-$ interactions 
for $E_\nu \simeq 6.3 \times 10^{15}$ eV---we adopt the detection 
probability set out by 
Anchordoqui et~al. (2004a)
%\citet{Anchordoqui2004b} 
with their specification of 
an effective target volume of $\sim 2$ km$^3$ for the IceCube
detector (which we also adopt for the hypothesized
Mediterranean detector).

Note that we assume in this work that a detector
can perfectly determine the energy of the primary neutrino.
This is a good approximation for our purpose
which is simply to determine
the observability of
the GC neutrino flux.

For $\hbox{Area}[E_\nu,\theta]$, in the case of IceCube we
employ the results of the Monte Carlo modeling
presented by %\citet{Ahrens2004}
Ahrens et al.~ (2004a)
and for a future
Mediterranean detector we assume an energy- and nadir-angle-dependent
fiducial (muon) area of 1 km$^2$.

\begin{figure}
\epsscale{1.0}
\plotone{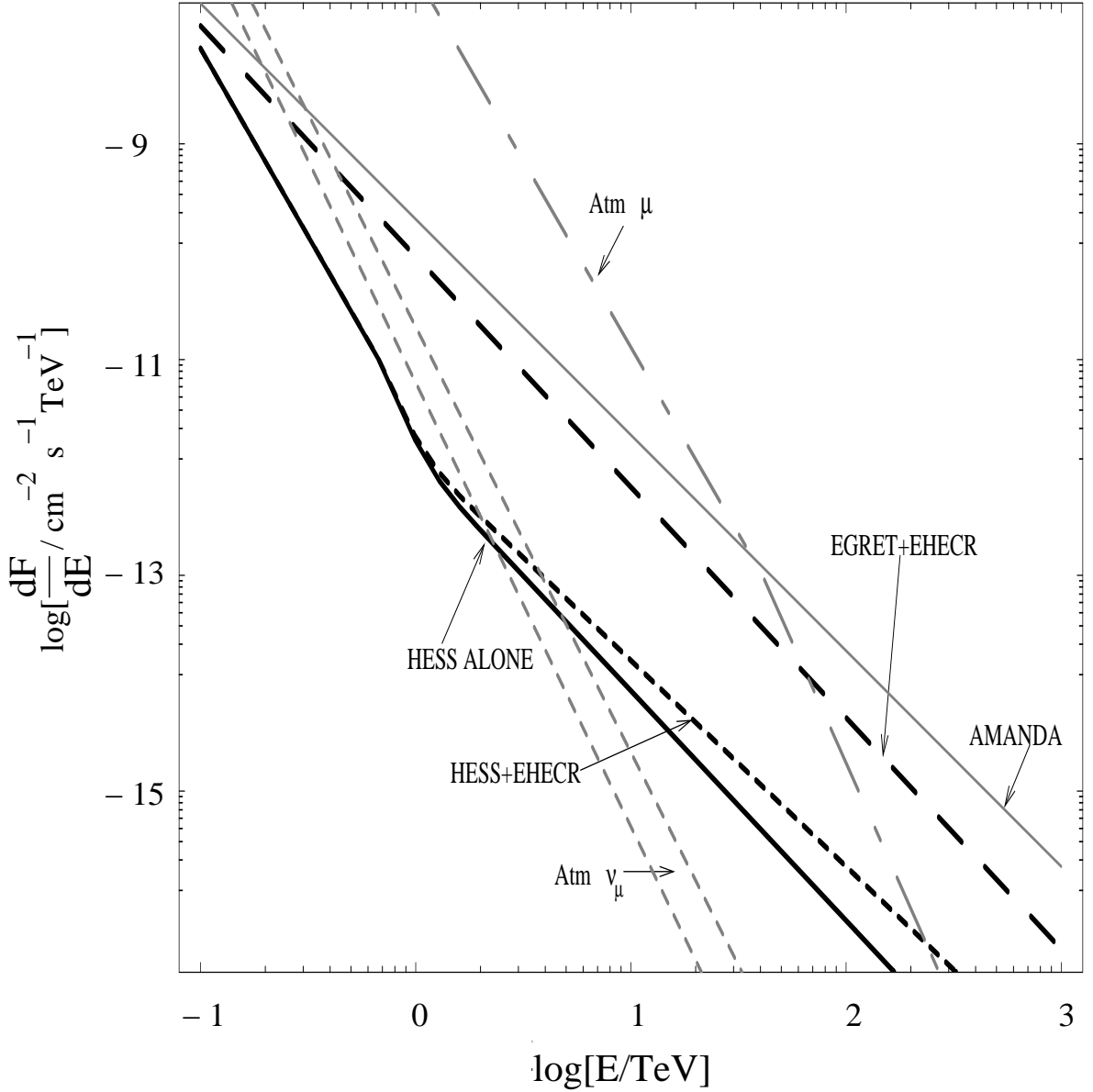}
\caption{GC neutrino fluxes.
Signals (black curves) with normalizations: 
(i) EGRET+EHECR---long dashed,
(ii) EGRET with cut-off (at 200 GeV) combined with HESS+EHECR---short dashed, 
and (iii)
EGRET with cut-off (at 200 GeV) combined with HESS ALONE
(with $10^{17}$ eV cut-off)---unbroken. The two dashed
gray curves give the upper and lower limiting values of the atmospheric
neutrino background in a km$^3$ Mediterranean detector and
the single gray dot-dashed curve gives the atmospheric muon
background in IceCube.
For reference,
the solid gray curve gives the present AMANDA limit on a neutrino source
in the {\it northern} sky \citep{Ahrens2004b} with $E^{-2}$ spectrum.
\label{figl}}
\end{figure}

The background to the CC muon production  process
is generated by the atmospheric muon (for an above the horizon source)
and neutrino fluxes \citep{Chirkin2004} within the solid angle
defined by the detector angular resolution. 
The background to the showering
and Glashow processes is due to the atmospheric neutrinos alone.
In IceCube a
0.7$^\circ$ angular resolution for a muon track
with $E_\mu \gtrsim 10^{12}$ eV 
is predicted (Ahrens et al. 2004, Anchordoqui et al. 2004a)
allowing for a search window of 1$^\circ \times 1^\circ$. 
Note that, as the GC is always overhead from the South Pole,
its CC \numu \ signal 
(i) is invisible to AMANDA \citep{Ahrens2004b}
and (ii) can only be seen
above $\sim$10$^{14}$ eV
in IceCube
(given the atmospheric muon background
at a fiducial depth of $\sim$1.6 km depth in the ice).
Further, Monte Carlo modeling by the IceCube collaboration shows that 
the detector effective (muon) area is significantly reduced
for a down-going neutrino flux at energies $\lesssim 10^{15}$ eV\citep{Ahrens2004}.
At higher energies, the effective area recovers, however,
and, further, the GC neutrino flux is not shadowed by the Earth in IceCube.
%Given these facts, 
Our analysis then indicates the GC should be detectable
in \numu \ CC events
by IceCube for both the EGRET+EHECR and HESS+EHECR cases 
(in the latter case, because
of the hard spectrum, the IceCube
event rate is actually comparable to that in a Mediterranean 
km$^3$ detector despite the necessarily higher energy threshold): see Figure \ref{figl}.

Relative to IceCube, a Mediterranean km$^3$ detector 
would have---in  the GC
CC \numu \ signal context---the advantage of both 
a largely
below-horizon source and
a tighter angular resolution (due to the relatively
longer scattering length of \cerenkov \ light
in deep sea water in comparison with Antarctic ice).
This would allow such a detector, according to our 
calculations\footnote{These calculations
employ a parameterization
of the modeled angular resolution of 
ANTARES \cite{Korolkova2004}. A km$^3$-scale detector will at least match this.},
to detect the GC at energies $\sim$TeV
(see Figure \ref{figl}).

For showers the angular determination
is much worse than for muons---we assume 25$^\circ$
as determined by Beacom et ~al. (2003). We also assume the same
angular resolution for events due to the Glashow process.
%which will be
%indistinguishable from ordinary showers within the energy range
%$10^{6.7} \to 10^{6.9}$ GeV \citep{Anchordoqui2004b}.
%In contrast to the CC \numu \ case, 
For shower processes, IceCube certainly has the advantage over a 
Mediterranean detector in the GC context: atmospheric
muons do not significantly pollute the shower signal,  so
the imposition
of the Earth between source and detector only serves to attenuate the signal.

Other potential signals in a km$^3$ detector from the GC neutrino flux
are double-bang and lollipop events due to the CC interactions of
higher-energy
\nutau 's. Unfortunately, employing an event rate parameterization 
for these two
processes that follows from the work of 
%\citet{Beacom2003}, 
Beacom et~al. (2003)
we find
an undetectably small GC \nutau \ signal for all cases.
(even in 
the EGRET+EHECR case the rate of either double-bang lollipop 
events is less than 0.03 per year).
We have also checked whether the very high energy
component of the signal from the GC neutrino source
might be uncovered using ``alternative'' astrophysical
neutrino detection techniques relying on, e.g., 
 horizontal shower detection in
the Auger
cosmic ray air shower array  \citep{Bertou2002},  
%purpose-built 
%neutrino telescope arrays in the lee of mountains \citep{Hou2002}, 
or the RICE \citep{Kravchenko2003}
in-ice radio \cerenkov \ detector.
Unfortunately,
we find for every case investigated, a negligibly small signal.
%(for both Auger and RICE this follows substantially from geometry 
%considerations). 

Yearly event rates and backgrounds are displayed
in table
\ref{table_eventrates_EGRETEHECR}. This also shows
the expectation for the
number of years
required before a positive signal can be uncovered given the background
(in the narrow sense of requiring that
one achieves 95\% confidence
level assuming Poisson statistics -- it should be emphasized here that, pragmatically,
more than one neutrino detection and a higher level of statistical significance
would be required before a reasonable announcement of a detection could be made).
We have presented the signal event rate above an energy which guarantees
it is equal to or surpasses the background event rate. Detector-dependent
modeling would be required to optimize this threshold energy.
The CC \numu \ event rates for up and down-going
neutrinos in a Mediterranean detector are given separately
 because a sea-water based detector
might be restricted to purely up-coming events by
design considerations. For the EGRET+EHECR case we present the
CC \numu \ and shower event rates in IceCube above two threshold energies
(design considerations may mean that IceCube only
attains $4 \pi$ sensitivity above $10^{14}$ eV).
In the case
of the HESS ALONE \numu \ (up) signal, we present the event rate for two different
values of the cut-off energy in the HESS source spectrum
as previously explained [though note that
for both \numu \ CC in IceCube and \numu \ CC (down) in Med km$^3$
cases we set $E_\gamma^{cut} = 10^{17}$ eV, 
the lower-energy cut-off being undetectable in those cases].

\section{Conclusion}
We have calculated the expected flux of neutrinos from the GC given the high-energy,
astroparticle signals that have been detected from this region.  From these flux 
estimates we have predicted event rates in three neutrino telescopes: IceCube, 
ANTARES, and a km$^3$-scale successor to ANTARES.  Recent data from HESS 
mean that earlier estimates for GC neutrino fluxes are likely to be
over-optimistic, though the possibility that \gam -\gam \ attenuation is reducing 
the $\sim$TeV gamma-ray flux means that such high fluxes are not excluded. 
In this most optimistic case the GC would  be seen within a year by ANTARES.
Even if the \gam -\gam \ attenuation is not operating, the HESS data
together with the EHE cosmic ray data on the GC now strongly suggest 
an interesting signal for {\it both} a Mediterranean km$^3$ detector and
IceCube at the South Pole.  This signal, in the most likely scenario, 
would be detected within about 1.5 years
by either IceCube or a Mediterranean km$^3$ detector. 
%\null\vskip-0.5in\null
\section{Acknowledgments}
This research was supported in part by NASA grant NAG5-9205 and 
NSF grant AST-0402502 at
the University of Arizona. RRV and FM are also partially supported by a joint
Linkage-International grant from the Australian Research Council.
RMC thanks Luis Anchordoqui, Paolo Deisati, Francis Halzen,
Gary Hill, Karl Kosack, and Josh Winn for useful correspondence.
%\null\vskip-0.5in\null

\clearpage

\begin{table}[h]
\begin{tabular}{|l|llllll|} 
\hline
 & dtctr & prcss & rate & thrsh & bkgd & yrs\\
\hline
\hline
E&ICE & \numu  & 9.2 &13.5 & 0.8  & 0.20 \\
G&        &     & 7.1 & 14.0 & 0.04  & 0.14 \\
R&& shwr               &  31 & 13.5 & 7  & 0.053 \\
E&&&  13 &14.0 & 0.5  & 0.074 \\
T&& $\overline{\nu}_e$-rsnt & 1.9 &&  0.2  & 0.95 \\
\cline{2-7}
+&ANT &  \numu  & 1.0 &11.7 &  0.02   & 0.98 \\
E&& shwr    &  0.2 &13.5 & 0.1  & 13 \\
\cline{2-7}
H&MED &  \numu  (up)& 101.0 &11.7 & 6.1   & 0.019 \\
E&&  \numu  (dn)& 14.1 &13.0 & 0.2  & 0.070 \\
C&& shwr  &  21.6 &13.5 & 10.0  & 0.063 \\
R&& $\overline{\nu}_e$-rsnt & 1.9& & 0.2   & 0.95 \\
\hline
\hline
H& ICE  & \numu  & 0.6 &14.1 & 0.02  & 1.6 \\
E  +&& shwr    &  0.8 &14.0 & 0.5 & 3.9 \\
S \ E&& $\overline{\nu}_e$-rsnt & 0.3& & 0.2  & 10 \\
\cline{2-7}
S \ H&MED & \numu  (up)& 2.0 &12.0 & 1.3  & 1.5 \\
\ \ \ \ E&& \numu   (dn)& 0.7 &13.0 & 0.2  & 3.3 \\
\ \ \ \ C&& shwr   &  0.3 &14.0 & 0.2  & 10 \\
\ \ \ \ R&& $\overline{\nu}_e$-rsnt & 0.3& & 0.2 & 10 \\
\hline
\hline
H \ A& ICE & \numu  & 0.1 &14 & 0.04 & 21 \\
\cline{2-7}
E \ L&MED & \numu  (up) &&&& \\
S \ 0& $(E_\gamma^{cut}=$ &$ 10^{17}$ eV)  & 0.9 &12.3  & 0.3 & 2.5 \\
S \ N& $(E_\gamma^{cut}=$ &$ 10^{13}$ eV) & 0.3 &12.3  & 0.3 & 12 \\
\ \ \ \ E& & \numu (dn) & 0.1 &13.5 & 0.006 & 19 \\
\hline
\end{tabular}   
\caption{Yearly event rates (``rate'') and backgrounds (``bkgd'') due to the various neutrino 
interaction processes (``prcss'') and normalizations specified
(`ICE', `ANT', and `MED' denote events in IceCube, ANTARES and a km$^3$ Mediterranean detector respectively). 
Backgrounds are over the same energy range as observations (above
a threshold energy,
``thrsh'', which is specified by log[$E_{th}$/eV] ).
Also displayed (``yrs'') is
the expectation for the 
number of years required before a real signal can be 
uncovered at the 95\% confidence level.
\label{table_eventrates_EGRETEHECR}
}
\end{table}

\clearpage

\end{document}